\documentclass[aps,showpacs,showkeys,preprintnumbers,preprint,groupedaddress]
{revtex4}

\usepackage{graphicx}

\begin{document}

\preprint{Physics Letters A}

% front/main.tex

\title{The Gap-Tooth Method in Particle Simulations}

\author{C. William Gear$^{1}$, Ju Li$^2$, Ioannis G. Kevrekidis$^3$}

\affiliation{$^1$NEC Research Institute,
4 Independence Way, Princeton, New Jersey 08540}

\affiliation{$^2$Department of Materials Science and Engineering, Ohio
State University, Columbus, Ohio 43210}

\affiliation{$^3$Department of Chemical Engineering, PACM and
Mathematics, Princeton University, Princeton, New Jersey 08544}
\email{yannis@Princeton.EDU}

\date{\today}

\begin{abstract}
We explore the gap-tooth method for multiscale modeling of systems
represented by microscopic physics-based simulators, when
coarse-grained evolution equations are not available in closed form.
A biased random walk particle simulation, motivated by the viscous
Burgers equation, serves as an example.
We construct macro-to-micro (lifting) and micro-to-macro (restriction)
operators, and drive the coarse time-evolution by particle simulations
in appropriately coupled microdomains (``teeth") separated by large
spatial gaps.
A macroscopically interpolative mechanism for communication between
the teeth at the particle level is introduced.
The results demonstrate the feasibility of a ``closure-on-demand''
approach to solving hydrodynamics problems.
\end{abstract}

\pacs{02.70.-c, 47.11.+j}

\keywords{modeling; multiscale; closure-on-demand;
gap-tooth; lifting; restriction}

\maketitle

% introduction/main.tex

Traditional approaches to solving physical problems that manifest
separation of scales involve first (a) deriving a set of reduced
equations to describe the system, and subsequently (b) solving the
equations and analyzing their solutions.
Recently an ``equation-free'' approach has been proposed
\cite{Kevrekidis02} that sidesteps the necessity of first deriving
explicit reduced equations.
The approach relies instead on microscopic simulations, enabling them
through a computational superstructure to perform numerical tasks {\it
as if the reduced equations were available in closed form}.
Both macroscopically coarse-grained equations and atomistic/stochastic
simulations can be regarded as ``black boxes" from the point of view
of appropriately formulated numerical algorithms.
They constitute alternative realizations of the same macroscopic
input-output mapping.
For example, a crystal's elastic response can either be {\it
evaluated} using elastic constants, or {\it estimated} by a
high-accuracy electronic structure program based on density functional
theory, which, for a given strain, computes the stress on-the-fly.
The advantage of a simulator-based approach is that it can be used
generally, beyond the region of validity of any given closure -
e.g. providing the correct nonlinear elastic responses in the above
example.
Equation-free methods hold the promise of combining direct
physics-based simulation with the strength and scope of traditional
numerical analysis on {\em coarse} variables (bifurcation, parametric
study, optimization) for certain problems - problems for which coarse
equations conceptually exist, but are not available in closed form.
An example is the so-called interatomic potential finite-element
method (IPFEM)\cite{Li02}, a subset of the more general
quasi-continuum method\cite{Phillips99}, used to identify elastic
instabilities leading to defect nucleation in nanoindentation, for
which no accurate closed-form constitutive relation is currently
available due to the complex triaxial stress state at the critical
site of instability.

Microscopic simulations cannot be used directly to attack problems
with large spatial and temporal scales (``macrodomains" in space and
time); the amount of computation is prohibitive.
If, however, the actual behavior can be meaningfully coarse-grained to
a representation that is smooth over the macrodomain, the microscopic
systems need only be directly simulated in {\it small patches} of the
macrodomain.
This is done by interpolating hydrodynamic variables between the
patches in space - the gap-tooth method (see \cite{Kevrekidis00}) -
and extrapolating from one or more patches in time - projective
integration\cite{Gear01a,Gear02a}.
In this paper, we use this ``closure-on-demand'' approach to solve for
the coarse-grained behavior of a particular microscopic system.
The illustrative example is the biased random walk of an ensemble of
particles, motivated by the viscous Burgers equation,
\begin{equation}
 u_t + uu_x = \nu u_{xx},
 \label{BurgersEquation}
\end{equation}
a 1D version of the hydrodynamics equations used under various
conditions to model boundary layer behavior, shock formation,
turbulence, and transport.
Here, $\nu>0$ is the viscosity; periodic boundary condition is used
for simplicity, and only non-negative solutions $u(x,t)>0$ are
considered.
A particular microscopic dynamics is constructed, motivated by
Eq.(\ref{BurgersEquation}), interpreting $u$ as the {\it density
field} of the random walkers; $\int udx=1$ corresponds to $Z$ walkers,
where $Z$ is a large normalization constant.
In the micro-simulation, random walkers move on $[-1,1)$ at discrete
timesteps $t_n=nh$. At each step, an approximation to the local
density, $\rho_i$, is computed from the positions of neighbors.
Then every walker is moved by $\Delta x_i \in N(h\rho_i/2,2\nu h)$, a
biased Gaussian distribution.
$x_i$'s are then wrapped around to $[-1,1)$, and the process repeats.
Since $\rho_i$ is a local estimate of $u$, this process achieves a
coarse-grained flux analogous to $j\equiv u^2/2-\nu u_x$ in
Eq.(\ref{BurgersEquation}) by assigning each walker a drift velocity
of $\rho_i/2$.

% In the following section we are going to discuss the gap-tooth scheme %
% which partitions space into regions (the teeth) in which the %
% microscopic simulation is carried out, and the gaps over which the %
% solution is interpolated. %

% gaptooth/main.tex

% \section{The Gap-Tooth Scheme} %

\begin{figure}
%\begin{figure}[ht]
\centerline{\includegraphics[scale=0.5]{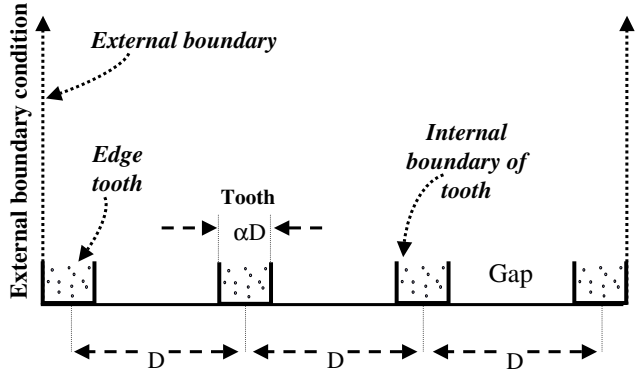}} \caption{Teeth
and gaps covering space.} \label{s1}
\end{figure}

The gap-tooth scheme, first discussed in \cite{Kevrekidis00}, covers
space with teeth and intervening gaps as shown in Fig. \ref{s1} for
one dimension.
The microscopic evolution is  simulated in the interior of
each tooth.
Clearly appropriate boundary conditions have to be provided at the
edges of each tooth.  Tooth boundaries coincident with external
boundaries have the boundary conditions specified externally, while
internal boundary conditions must be generated by the gap-tooth scheme
itself.  Because this example uses periodic boundary conditions, there
are no external boundaries: the teeth can be viewed as equally spaced
on a circle.

The microscopic simulation operates on the position of each particle.
We are interested in a meaningful {\it coarse} description, a
finite-dimensional approximation to the density of particles, $u(x)$,
possibly averaged over several realizations of the computational
experiment \cite{Kevrekidis02}.
The {\it lifting} operator that maps a given $u(x)$ to consistent
particle positions is straightforward in this case.
From the density function over a tooth we can compute its integral,
so we know the number of particles that should be present in that
tooth.
The indefinite integral of the density function over the tooth
provides the cumulative distribution function for that tooth which
permits the particles to be placed as a discrete representation of
that function\cite{cwgdist}.
If the density approximation is constant
in each tooth (as has been found to be adequate in the examples here)
this simply means that the particles are uniformly spaced in each
tooth according to the density in that tooth.

In our particular stochastic simulation, the evolution rules require a
local density estimate.
This should be done by choosing a particle density influence function
$\sigma(d)$ that specifies the contribution of each particle to the
local density at a distance $d$ from that particle.
If this function is constant for $d < w$ and zero elsewhere, the local
density function at $x$ is found by just counting the number of
particles within distance $d$ of $x$.
Ref. \cite{Chertock01,Chertock02}, seeking some level of
differentiability, use a Gaussian spreading function for each
particle.
Since we do not require differentiability, we will count particles
within distance $d$.
By making $d$ twice the tooth size, we can then simply count the
particles in each tooth.
We have also used higher-order approximations, but it is not clear
that they yield sufficient improvements in accuracy to justify the
additional computational effort.
The technique used for higher-order approximations is based on the
fact that the sample cumulative distribution function in each tooth is
known from the particle positions.
We can then fit a polynomial to it within each tooth.
The density function over a tooth is simply given by the derivative of
this polynomial.

The mapping of a phase point or points (particle positions and
velocities history) to coarse fields is called a {\it restriction}
operator. In addition to the density field ($0$th-moment), smooth
velocity ($1$st-moment) and temperature ($2$nd-moment) fields can be
extracted from molecular dynamics based on maximum likelihood
inference \cite{Li98}. If the interior of a tooth were to be simulated
by solving a PDE, we would need to prescribe appropriate boundary
conditions at each tooth at each timestep. The same is still true when
the tooth is realized using particle-based microscopic
simulations. Creating an appropriate match between the coarse fields
at the boundaries and the particles in the teeth is an area of intense
research\cite{Li99,E01}.  Sometimes one knows so little about the
nature of the coarse equation that even the correct {\em order} for
imposing well-posed boundary conditions at the teeth is unknown. This
issue is addressed in \cite{Li03}.

% boundary/main.tex

% \subsection{Particle Boundary Conditions} %

Here we use an alternative approach suggested in \cite{Gear02b}.  In a
1D particle based random walk simulation we can distinguish two
``fluxes'' - left-going and right-going.
The particle simulation in the interior
of each tooth generates {\it  outgoing} fluxes,  that is, the
left[right]-going fluxes at
the left[right] boundaries,  directly.
Boundary conditions are needed to
provide matching {\it incoming} (right[left]-going) fluxes at the same
boundaries.
In $d$-dimensions, there will be $2^d$ boundaries to deal with and the
corresponding incoming fluxes to provide.

\begin{figure}
\centerline{\includegraphics[scale=0.55]{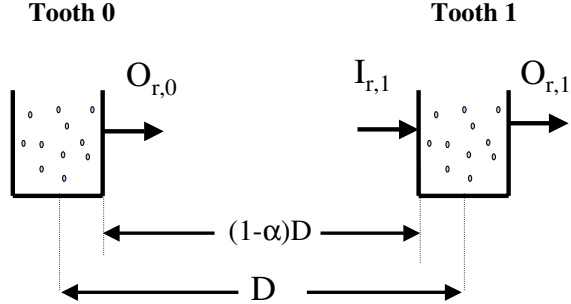}}
\caption{Right-going Input and Ouput Fluxes} \label{s2}
\end{figure}

Consider the estimation of the right-going incoming flux, ${\rm
I}_{r,1}$, as shown in Fig. \ref{s2}.
Assuming macroscopic flux smoothness suggests that we can interpolate
its values from neighboring {\it outgoing} fluxes, in this case ${\rm
O}_{r,0}$ and ${\rm O}_{r,1}$.
If we use linear interpolation, we can write
\begin{equation}
{\rm I}_{r,i} = \alpha {\rm O}_{r,i-1} + (1 - \alpha){\rm O}_{r,i}.
\label{linear}
\end{equation}
The interpolation coefficients depend (in this case through $\alpha$)
only on the gap-tooth geometry.

\begin{figure}
\centerline{\includegraphics[scale=0.55]{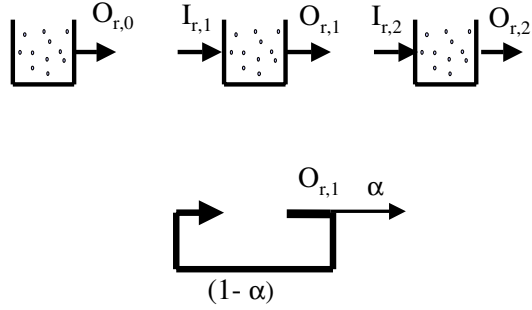}} \caption{Flux
Redistribution for Right-going Fluxes} \label{s3}
\end{figure}

However, the ``fluxes'' under discussion here are not real-valued
quantities, but discrete events as particles cross a boundary, so
Eq.(\ref{linear}) needs a different interpretation.
Consider instead the role played by each {\it outgoing} flux in the
interpolation for {\it ingoing} fluxes.
An interpretation of Eq.(\ref{linear}) for $i = 1$ and $i = 2$ would
be the portion $(1 - \alpha)$ of ${\rm O}_{r,1}$ contributes to the
flux ${\rm I}_{r,1}$ while $\alpha$ of it contributes to ${\rm
I}_{r,2}$.
A similar procedure applies to the left-going fluxes.
Thus, rather than thinking in terms of flux interpolation we can think
in terms of {\em flux redistribution}.
Interpreting the linear interpolation stochastically, (on a regularly
spaced gap-tooth scheme) we direct $\alpha$ of the outgoing particles
as input to the neighboring tooth, and redirect $(1-\alpha)$ of them
back to the other boundary of the same tooth as shown in
Fig. \ref{s3}.

Flux redistribution has to recognize the position of a particle after
it leaves a tooth.
If it had moved to a distance $\delta$ beyond the boundary of the
tooth, it must be inserted a distance $\delta$ inside the receiving
tooth.
If $\delta$ were larger that the tooth width it would have left a
tooth boundary again and a further redistribution would be required
following the same rule.
(In multiple dimensions, the boundaries in each dimension are treated
independently so that a particle will be distributed for each boundary
that it crosses until it is inside a tooth.)

The above method implements effective linear interpolation.
As discussed in \cite{Gear02b}, linear interpolation is not adequate
for second-order problems: at least quadratic interpolation must be
used.
A possible quadratic interpolation formula is
\begin{equation}
 {\rm I}_{r,i} = \frac{\alpha(1+\alpha)}{2} {\rm O}_{r,i-1}+(1 -
\alpha^2) {\rm O}_{r,i}-\frac{\alpha(1 - \alpha)}{2} {\rm O}_{r,i+1}.
\nonumber
\label{quad}
\end{equation}
As before we consider the impact of each outgoing flux on incoming
fluxes.
The fractions of output ${\rm O}_{r,1}$ should be sent to the inputs
as follows: $(1 - \alpha^2)$ to ${\rm I}_{r,1}$; $\alpha(1+\alpha)/2$
to ${\rm I}_{r,2}$: and $-\alpha(1-\alpha)/2$ to ${\rm I}_{r,0}$.

Note that the last value is negative.  Any linear higher-order
interpolation formula contains negative coefficients.
Our solution is to direct {\em anti-particles} to the appropriate
teeth.
There they must annihilate with regular particles - we simply
annihilate with the nearest regular particle.
With this approach, the ${\rm O}_{r,1}$ is redistributed as follows:
$(1 - \alpha(1+\alpha)/2)$ to ${\rm I}_{r,1}$; $\alpha^2$ to ${\rm
I}_{r,2}$; and $\alpha(1-\alpha)/2$ are cloned to get two regular
particles sent to ${\rm I}_{r,1}$ and ${\rm O}_{r,2}$ and one
anti-particle sent to ${\rm I}_{r,0}$
It is interesting to observe that this scheme conserves the total
number of particles in the computation.

% burgers/main.tex

% \section{Coarse-grained Results} %

\begin{figure}
\centerline{\includegraphics[scale=0.4]{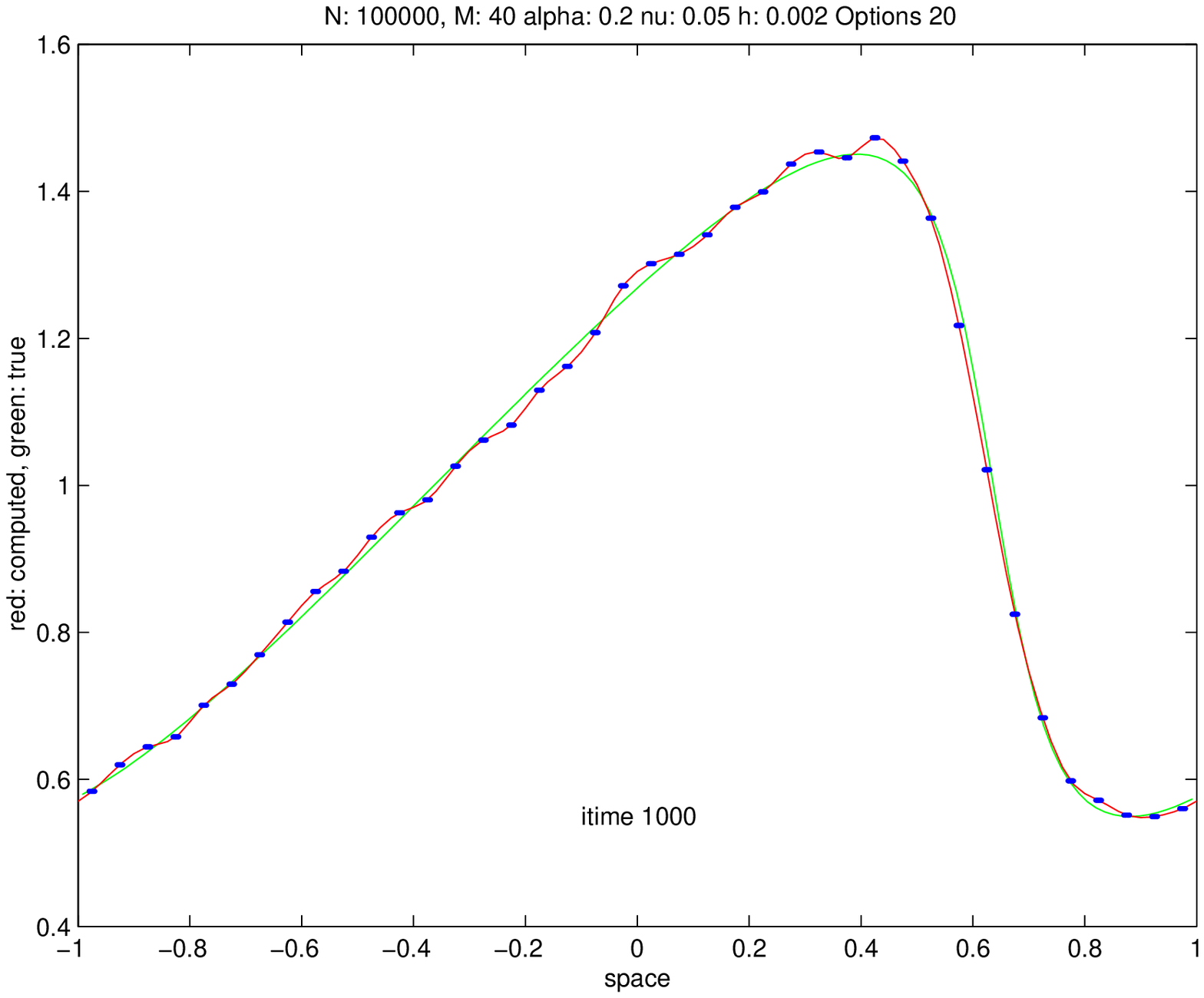}} (a)
\centerline{\includegraphics[scale=0.4]{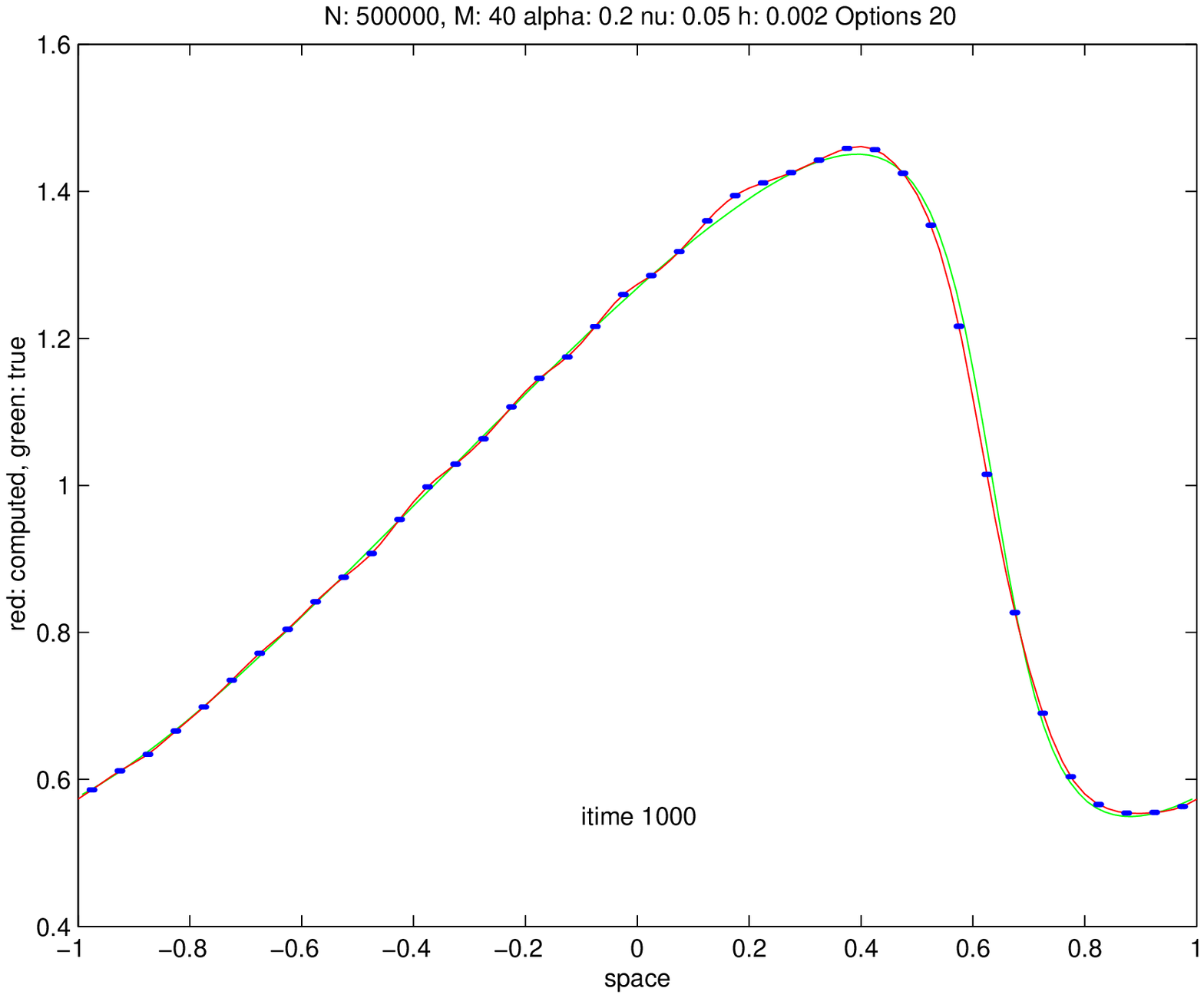}} (b)
\caption{Simulation results after 1,000 steps.  $\alpha = 0.2$.
(a) 100,000 particles (b) 500,000 particles} \label{fr1}
\end{figure}

The microscopic evolution rules were simulated at conditions
corresponding to $\nu = 0.05$ and timestep $h = 0.002$ in
Eq.(\ref{BurgersEquation}) for 1,000 time steps using the gap-tooth
scheme with 40 equally spaced teeth in the interval $[-1,1)$.
The tooth-to-spacing ratio was $\alpha = 0.2$.
The coarse initial condition was $1-\sin(\pi x)/2$.
Fig. \ref{fr1} shows the results at $t = 2$ when 100,000
particles are used (upper figure) and 500,000 particles (lower
figure).
The dots indicate the particle density approximations within each
tooth (constant in this case); they are connected by a spline
interpolant.
The smooth curve (to guide the eye) is the analytical solution
\cite{Cole51,Hopf50} of Eq.(\ref{BurgersEquation}).

\begin{figure}
\centerline{\includegraphics[scale=0.4]{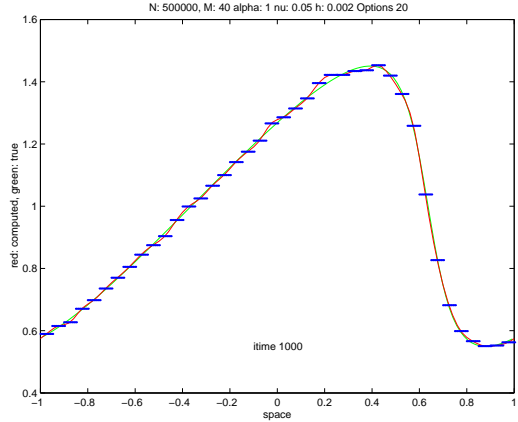}}
\caption{Standard particle simulation (no gaps)} \label{fr2}
\end{figure}
The problem was also run with $\alpha = 1$, i.e., no gaps, to compare
with a conventional, full space, direct microscopic simulation.
Fig. \ref{fr2} shows the result using 500,000 particles and a constant
density approximation in each tooth.
This is the same average particle density per spatial unit as with
$\alpha = 0.2$ and 100,000 points.
The results are comparable.
%
%One might  object that the constant density approximation over
%the larger teeth in this case is a cause of some of the error which
%would otherwise be smaller.
%
%Fig. \ref{fr3} shows the results from
%using 200 teeth with $\alpha = 1$, 500,000 particles, and constant
%density after just 10 time steps.
%
%It can be seen that this leads to
%much greater ``noise'' from the stochastic simulation.
%
%The effect of
%the particle redistribution system for handling boundary conditions is
%to reduce the relative number of particles that change teeth when
%$\alpha$ is smaller, and this serves to reduce the stochastic
%variation.
%
The $\alpha = 1$ case with 40 teeth and 500,000 particles has also
been run with the polynomial density interpolants described above of
degrees 1 and 2.
No significant difference in the solutions were observed, and neither
were differences observed when the higher order polynomial fits were
used with $\alpha = 0.2$.

%\begin{figure}
%\centerline{\includegraphics[scale=0.4]{burga1n50m200.ps}}
%\caption{No gaps, 200 teeth.}
%\label{fr3}
%\end{figure}

% comments/main.tex

% \section{Comments} %

We have demonstrated that the gap-tooth scheme can be successful in
solving some problems using microscopic models based on the stochastic
simulation of particle motion; we also introduced a novel approach for
dealing with the inter-tooth boundary conditions.

In earlier work we have proposed combining this with projective
integration \cite{Gear01a} and some preliminary experiments have been
performed in this direction.
However, projective integration requires smoothness of the time
derivative estimates.  The stochastic nature of the microscopic model
leads to significant noise.  In this simulation we used a
least-squares linear estimate from a large number of time steps to get
a reasonably accurate time derivative estimate.  However, by then, the
total time step at the microscopic level was large relative to the
size of a projective step in time (which is limited by the smoothness
of the solution).
If the stochastic noise is reduced by using a much larger number of
particles, or a large number of ``copies" of the simulation, the time
derivative estimates would allow the application of projective
integration.
In some sense there is a tradeoff between saving computation in the
spatial domain with fewer teeth and fewer particles, and saving
computation in the time domain by getting more accurate estimates of
the time derivatives.
We believe that for problems where there is a significant gap between
the timescales of the microscopic dynamics and those of the expected
macro-scale behavior, projective intregration would be a useful
addition.
We will report on such ``patch dynamics" experiments in a future paper.

% acknowledgment/main.tex

This work was partially supported by the Air Force Office of
Scientific Research (Dynamics and Control) and
an NSF ITR grant (C.W.G.,I.G.K.).
J.L. acknowledges support by Honda R\&D Co., Ltd. (RF\#
744252) and the OSU Transportation Research Endowment Program.

% bibliography/main.tex

\end{document}